# Phylo.io: interactive viewing and comparison of large phylogenetic trees on the web

Oscar Robinson[1*], David Dylus[2,3*], Christophe Dessimoz[1,2,3#]

[1]Department of Computer Science & Dept. of Genetics, Evolution and Environment, University College London, Gower Street, London WC1E 6BT, UK

[2]Dept. of Ecology and Evolution & Center for Integrative Genomics, University of Lausanne, Biophore, 1015 Lausanne, Switzerland

[3]Swiss Institute of Bioinformatics, Biophore, 1015 Lausanne, Switzerland

*Joint first authors.

[#]Corresponding author. Email: Christophe.Dessimoz@unil.ch

**ABSTRACT (150 Words)**

Phylogenetic trees are pervasively used to depict evolutionary relationships. Increasingly, researchers need to visualize large trees and compare multiple large trees inferred for the same set of taxa (reflecting uncertainty in the tree inference or genuine discordance among the loci analysed). Existing tree visualization tools are however not well suited to these tasks. In particular, side-by-side comparison of trees can prove challenging beyond a few dozen taxa. Here, we introduce *Phylo.io*, a web application to visualize and compare phylogenetic trees side-by-side. Its distinctive features are: highlighting of similarities and differences between two trees, automatic identification of the best matching rooting and leaf order, scalability to very large trees, high usability, multiplatform support via standard HTML5 implementation, and possibility to store and share visualisations. The tool can be freely accessed at http://phylo.io. The code for the associated JavaScript library is available at https://github.com/DessimozLab/phylo-io under an MIT open source license.

## INTRODUCTION

Phylogenetic analyses often require the inference and evaluation of multiple trees for the same group of taxa—either to gauge the uncertainty in the inference (e.g. by sampling trees from the posterior distribution) or to observe incongruence among different loci (e.g. resulting from horizontal gene transfer). Therefore, there is an increasing need for tools that facilitate quantitative and qualitative comparison of phylogenetic trees. In particular, the need for visualisation of large trees has driven the development of many different approaches, some of which are not only related to applications in biology (reviewed in Landesberger *et al.*, 2011).

Popular tools for tree visualisations are FigTree (Rambaut, 2009) and EvolView (Zhang *et al.*, 2012). FigTree allows users to display and manipulate tree visualisations in detail. Written in Java, it is a platform-independent standalone tool. Multiple trees can be loaded simultaneously and it is possible to browse through each individual tree. EvolView, on the other hand, is accessible through a web interface. EvolView allows users to map this information onto the tree visualisation, thus providing the user with an understanding of evolutionary events on the genome level. While both tools are well suited to analyse single trees, there is no function to compare different topologies. Moreover, visualisation of very large trees can become cumbersome. An additional tool to visualize phylogenetic trees is part of a bigger software package, *i.e.* the molecular evolutionary genetics analysis (MEGA) package (Tamura *et al.* 2013). This tool allows to display basic newick trees and to manipulate a single tree. However, similar to other tools, comparison of different trees is not supported.

Phylogenetic network visualisation tools such as SplitsTree (Huson and Bryant, 2006) and DensiTree (Bouckaert, 2010) display multiple trees at once. SplitsTree represents trees as a network and DensiTree overlays trees to create a composite image. However, the aggregated resulting visualisations make it difficult to pinpoint the specific changes between two trees. Furthermore, when the size of the trees increases, the legibility of the visualisations rapidly degrades.

A common method for side-by-side tree comparison is the use of “tanglegrams” (Scornavacca *et al.*, 2011). In this approach, leaves corresponding to the same species are linked with lines, with the dissimilarity between the two trees reflected in the number of line crosses. However, tanglegrams are difficult to read and interpret when there are substantial differences between two trees, and they too scale poorly to large trees.

For direct tree comparison, the best tools rely on colour coding and visual node annotation. Compare2Trees (Nye *et al.*, 2006) annotates nodes using a novel algorithm for tree structure comparison. However, it does not scale well when large trees are compared. Although not specifically developed to visualize phylogenetic trees, TreeVersity (Gomez *et* *al.*, 2012) is another tool that facilitates detailed visual comparison of trees, but is again limited in terms of its scalability. Another tool for side-by-side comparison is TreeJuxtaposer (Munzer *et al.*, 2003). It is scalable for very large trees and enables users to compare subtrees. Unfortunately, it has not been maintained for a long time and is difficult to run.

The final drawback with some current state of the art tools is their availability. Compare2Trees, a web based tool, for example is build as Java Applet and therefore difficult to use on modern web browsers. Desktop software, on the other hand, may require legacy systems or only run on specific platforms. Significant contributions such as PhyloComp (Bremm *et al.*, 2011) are, unfortunately, not readily available for download and installation. Moreover, PhyloComp also suffers from the same visual and computation scalability issues as many other tools. However, software facilitating phylogenetic tree visualisation in modern browsers are beginning to appear. One example is the jsPhyloSVG library (Smits and Ouverney, 2010). This library enables users to visualize and manipulate single trees. However, similarly to most other tools, it does not offer any function for comparing trees.

In summary, we can identify three main drawbacks with current tools: (i) a lack of functionality to compare trees; (ii) a failure to scale beyond a few dozen taxa; and/or (iii) the use of outdated technology, thereby compromising their usability.

Here, we present *Phylo.io*, a web-based software for tree viewing and side-by-side comparison. Similarities and differences between two trees are indicated with a colour scheme, and it is possible to highlight corresponding nodes and clades of interest between trees. The tool is scalable to large trees while maintaining tree legibility and quantitative comparison of the tree structures. Moreover, it is able to find the best corresponding rooting and order of leaves between two trees, facilitating their side-by-side comparison. Finally, the tool supports saving and sharing of a current state of visualisation via custom URLs, thereby facilitating collaborative work on large trees.

## NEW APPROACH

*Phylo.io* is built using HTML5, CSS, Ajax, jQuery and the D3 JavaScript visualisation library (Bostock *et al.*, 2011). Therefore, it does not need to be installed and is instantly accessible and usable on all modern web browsers. Additionally, all computations are performed on client side, making it inherently scalable.

In order to perform a tree analysis, the user can choose between two modes using the newick format as input. First, the “view” mode makes it possible to display a single tree. There, basic tree operations can be performed directly on the nodes and branches of the tree, *i.e.* re-rooting and branch swapping. Second, with the “compare” mode, two trees are displayed side by side. The differences and similarities of both trees are highlighted on the branches and nodes using a ColorBrewer colour scheme (Harrower and Brewer, 2003). The degree of similarity, indicated by a colour scale, is calculated using a variation of the Jaccard index that is optimized for speed, as presented in (Munzner *et al*. 2003; Bremm *et al*. 2011). All basic tree operations remain possible in this mode.

*Phylo.io* improves the legibility of large trees by automatically collapsing nodes so that an overview of the the tree remains visible at any given time. The underlying data structure stores tree nodes in an object that contains two lists, one with the visible nodes and the other with nodes that are collapsed. Therefore, if the rendering function reaches a node with collapsed subtree simply stops to render, thus making the interface more responsive. This also ensures that all data remain available for analysis but are not shown until required. During a first rendering pass, nodes beyond a certain depth (automatically estimated from the available screen size) are collapsed into a composite node that is represented by a triangle. The initial collapsing provides two benefits: (i) the tree remains legible, as branches are not bunched up and text is not overlapping; and (ii) the rendering is fast because not all subtree comparisons have to be calculated. Therefore, smaller trees are rendered instantaneously and larger trees in a relatively short time (for a tree with 500 taxa, a few seconds on a laptop). Furthermore, the tool provides a search functionality that colours the branches from a root to a queried leaf. In the “compare” mode, as branches are expanded, the best corresponding node in the opposing tree for each node in each newly visible branch is calculated. Therefore, the computationally expensive task of calculating the score and location of the best corresponding node is split up and only calculated just before the rendering. This enables the tool to remain as responsive as possible.

When comparing two trees, a common task is to match leaves in the tree or to find the best corresponding internal node in the opposing tree. In *Phylo.io*, the user can compute the best corresponding rooting. Moreover, an automated branch-swapping procedure allows users to find the best corresponding visualisation between two trees. This is useful in a broad range of contexts, such as to compare trees that have been inferred using different methods, to compare samples from a Bayesian posterior distribution, or to compare differences in the trees reconstructed from different loci.

For interactive analysis of specific parts of the tree, the user can select a node and highlight it. The best corresponding node in the opposing tree is then highlighted and centered, allowing the user to interactively match structures within the compared trees. If the best corresponding node is in a collapsed subtree, all nodes on the path to that node will be expanded making the node visible. Therefore, regardless of the current collapsing depth, the user can always find the corresponding structure in the opposing tree without having to view the entire tree.

Finally, *Phylo.io* allows users to share tree visualisations using the GitHub Gist API (which supports storage and retrieval of data). *Phylo.io* stores the current tree data structure in an extended newick format, where information about the visualization are preserved as metadata and thus save the current visualisation state. The share functionality generates a unique URL that can be shared with collaborators, who in turn can retrieve and work on the tree in its current state.

To illustrate the usefulness of *Phylo.io*, two trees generated using different methods, one using PhyML (Guindon *et al.*, 2010) and one using RAxML (Stamatakis, 2014), were computed on data retrieved from the OMA Orthology Database (Altenhoff *et al.*, 2015)—specifically the Hierarchical group HOG:0152954.2aq.7d at Embryophyta level. These trees contain 737 proteins and are viewable in the application as the “Large Example Trees” and are displayed in Fig. 1. In this example, although the two programs return seemingly very different trees, *Phylo.io* makes it obvious that most differences are merely due to the different rooting and subtree ordering.

In conclusion, *Phylo.io* addresses the need for a usable and scalable tree viewer, with particularly useful features for side-by-side tree comparison. However, as with any new tool, there is a long list of future functionality ideas. For instance, we would like to support other types of input formats and extend side-by-side comparison to trees with partially overlapping leaf sets. Meanwhile, by releasing *Phylo.io* under a permissive open source license, we also encourage improvements and bug fixes by the broader community.

**Acknowledgements**

We thank the members of the Dessimoz Laboratory for their help. Especially, we thank Ivana Piližota and Kevin Gori for providing the trees used as examples. Additionally, we thank Clément Train and Alex Warwick Vesztrocy for useful discussions about the implementation.

**Funding**

OR was supported by a Summer Research Bursary from the UCL Department of Computer Science. DD and CD acknowledge Swiss National Science Foundation grant 150654 and UK BBSRC grant BB/M015009/1.

## REFERENCES

Altenhoff AM, Škunca N, Glover N, Train C-M, Sueki A, Piližota I, Gori K, Tomiczek B, Müller S, Redestig H, et al. 2015. The OMA orthology database in 2015: function predictions, better plant support, synteny view and other improvements. Nucleic Acids Res. 43:D240–D249.

Bostock M, Ogievetsky V, Heer J. 2011. $D^3$: Data-Driven Documents. IEEE Trans. Vis. Comput. Graph. 17:2301–2309.

Bouckaert RR. 2010. DensiTree: making sense of sets of phylogenetic trees. Bioinformatics 26:1372–1373.

Bremm S, Von Landesberger T, Hess M, Schreck T, Weil P, Hamacherk K. 2011. Interactive visual comparison of multiple trees. VAST 2011 - IEEE Conference on Visual Analytics Science and Technology 2011, Proceedings:31–40.

Gomez J a. G, Plaisant C, Shneiderman B, Buck-Coleman A. 2012. Interactive Visualizations for Comparing Two Trees With Structure and Node Value Changes. HCIL Tech Report:1–11.

Guindon S, Dufayard J-F, Lefort V, Anisimova M, Hordijk W, Gascuel O. 2010. New algorithms and methods to estimate maximum-likelihood phylogenies: assessing the performance of PhyML 3.0. Syst. Biol. 59:307–321.

Harrower M, Brewer CA. 2003. ColorBrewer.org: An Online Tool for Selecting Colour Schemes for Maps. Cartogr. J. 40:27–37.

Huson DH, Bryant D. 2006. Application of phylogenetic networks in evolutionary studies. Mol. Biol. Evol. 23:254–267.

von Landesberger T, Kuijper A, Schreck T, Kohlhammer J, van Wijk JJ, Fekete JD, Fellner DW. 2011. Visual Analysis of Large Graphs: State-of-the-Art and Future Research Challenges. Comput. Graph. Forum 30:1719–1749.

Munzner T, Guimbretière F, Tasiran S, Zhang L, Zhou Y. 2003. TreeJuxtaposer. ACM SIGGRAPH 2003 Papers on - SIGGRAPH ’03:453.

Nye TMW, Liò P, Gilks WR. 2006. A novel algorithm and web-based tool for comparing two alternative phylogenetic trees. Bioinformatics 22:117–119.

Rambaut A. 2009. FigTree. Available from: http://tree.bio.ed.ac.uk/software/figtree/

Scornavacca C, Zickmann F, Huson DH. 2011. Tanglegrams for rooted phylogenetic trees and networks. Bioinformatics 27:i248–i256.

Smits S a., Ouverney CC. 2010. jsPhyloSVG: A javascript library for visualizing interactive and vector-based phylogenetic trees on the web. PLoS One 5:6–9.

Stamatakis A. 2014. RAxML version 8: a tool for phylogenetic analysis and post-analysis of large phylogenies. Bioinformatics 30:1312–1313.

Tamura K, Stecher G, Peterson D, Filipski A, Kumar S. 2013. MEGA6: Molecular Evolutionary Genetics Analysis version 6.0. Mol. Biol. Evol. 30:2725–2729.

Zhang H, Gao S, Lercher MJ, Hu S, Chen WH. 2012. EvolView, an online tool for visualizing, annotating and managing phylogenetic trees. Nucleic Acids Res. 40:569–572.

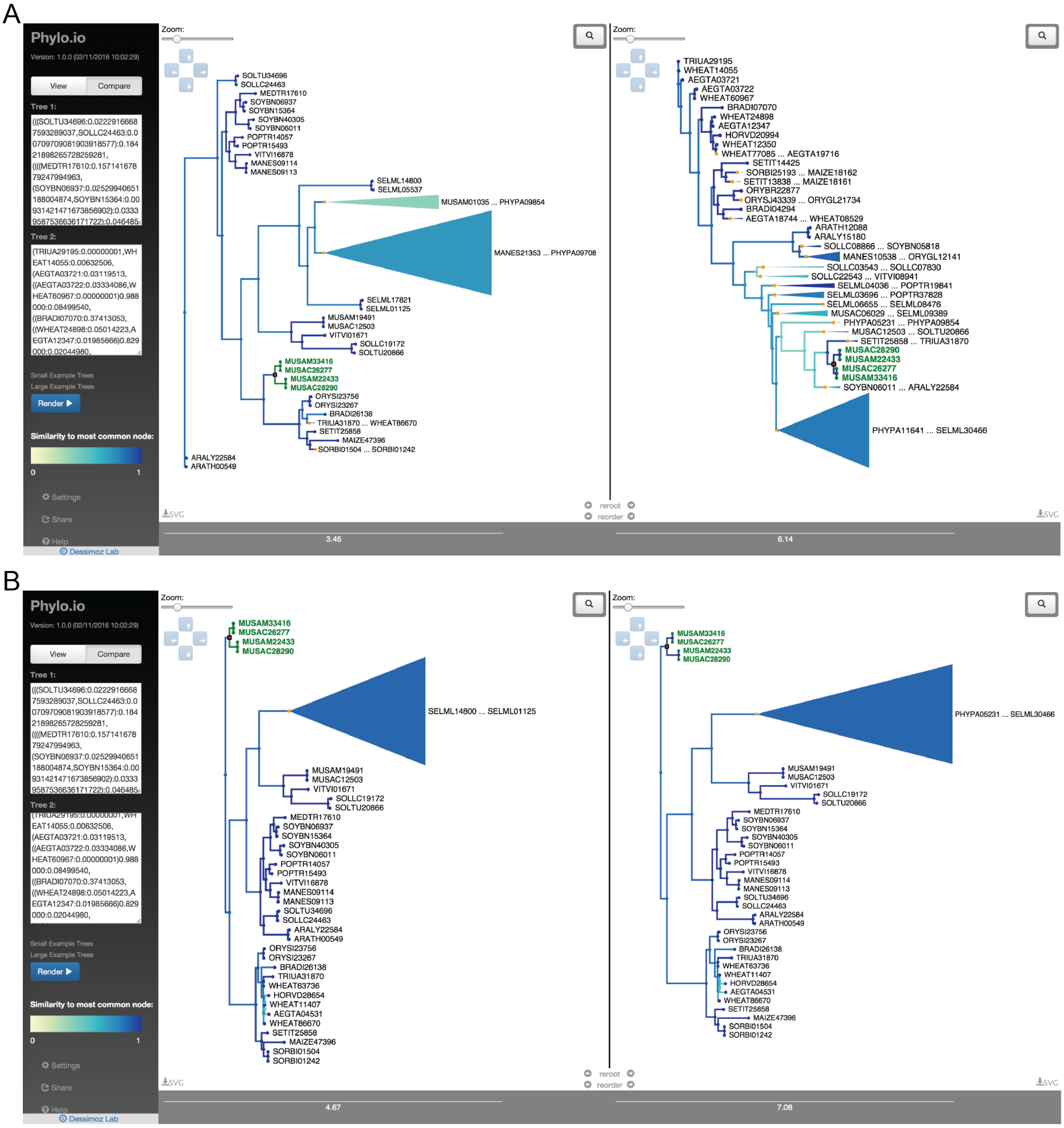

Figure 1. ***Phylo.io* web interface in "compare" mode**. (A) Two large trees with the same 737 leaves but different topologies in compare mode. The yellow to blue colour scheme indicates the similarity of best matching subtrees between the two trees. In the left tree, an inner node (highlighted in red) is selected, thereby highlighting its subtree (in green) and corresponding parts in the right tree (in green). (B) The same tree with the same highlighted inner node, but after automatically rerooting and reordering subtrees.